\documentclass[prl,aps,superscriptaddress,amsmath,twocolumn,showkeys,10pt,a4paper,showpacs]{revtex4-1}
\usepackage{graphicx}
\usepackage[caption=false,subrefformat=parens,labelformat=parens]{subfig}
\usepackage{url}
\usepackage[colorlinks,linkcolor=blue,citecolor=magenta,urlcolor=blue]{hyperref}
\usepackage{xcolor}
\usepackage{txfonts}
\usepackage[top=2.15cm,bottom=2.7cm,left=1.755cm,right=1.755cm]{geometry}

\makeatletter
\g@addto@macro\bfseries{\boldmath}
\makeatother

\usepackage{color}

\begin{document}
\title{Emergence of coexisting percolating clusters in networks}
\author{Ali Faqeeh}\author{Sergey Melnik}
\affiliation{MACSI, Department of Mathematics \& Statistics, University of Limerick, Ireland}
\author{Pol Colomer{-}de{-}Sim{\'o}n}
\affiliation{Departament de F\'isica Fonamental, Universitat de Barcelona, Mart\'{\i} i Franqu\`es 1, 08028 Barcelona, Spain}
\author{James P. Gleeson}
\affiliation{MACSI, Department of Mathematics \& Statistics, University of Limerick, Ireland}

\begin{abstract}
It is commonly assumed in percolation theories that at most one percolating cluster can exist in a network. We introduce sausage-like networks (SLNs), an ensemble of synthetic modular networks in which more than one percolating cluster can appear.
We show that coexisting percolating clusters (CPCs) emerge in such networks due to limited mixing, i.e., a small number of interlinks between pairs of modules. We develop an approach called modular message passing (MMP) to describe and verify these observations. We demonstrate that the appearance of CPCs is an important source of inaccuracy in the previously introduced percolation theories, such as the message passing (MP) approach. Moreover, we show that the MMP theory improves significantly over the predictions of MP for percolation on synthetic networks with limited mixing and also on several real-world networks. These findings have important implications for understanding the robustness of networks and in quantifying epidemic outbreaks in the susceptible-infected-recovered (SIR) model of disease spread.
\end{abstract}

\maketitle

Percolation theories are among the most studied in network science~\cite{Newman10}, as well as in several other areas \cite{christensen05,stauffer1994}, providing insights for a broad range of applications such as robustness of a network to random failures or attacks \cite{Callaway00}, epidemics in contact processes \cite{Munoz10_PRL}, vaccination strategies \cite{Newman10}, neuronal avalanches \cite{Friedman_choas13}, and stability of gene regulatory networks \cite{Squires_prl12}. In the simplest case of bond (or site) percolation, a fraction $p$ of the links (nodes) are randomly chosen to be occupied and the rest of the links (nodes) are removed from the network \cite{Newman10}. The quantity of interest is $\mathit{\overline{S}}$: the expected fractional size of the giant component (GC) of the network, which in the limit of infinitely large networks is referred to as the percolating cluster (PC) of the network \cite{Newman01e,christensen05}. The size of the GC scales linearly with the network size while the fractional sizes of other clusters vanish in the limit of infinitely large networks.

Theoretical approaches and extensive numerical simulations play pivotal roles in understanding and describing the behavior of percolation processes on networks. The $p_k$ theory for bond percolation \cite{Melnik11,Newman10}, for example, can accurately describe the results of numerical simulations on configuration model \cite{Molloy95} networks using only the network degree distribution. On networks with degree-degree correlations, the accurate results are obtained using the so called $P(k,k')$ theory \cite{Vazquez03} which employs the joint degree distribution. The $P_{k,k'}^{i,i'}$ theory \cite{Melnik_chaos14} can provide a more accurate description of dynamics on modular networks, as it considers the joint degree distributions within and between modules. The message passing (MP) approach \cite{karrer_PRL14} provides more accurate results than the aforementioned theories as it uses the full information on the adjacency of individual nodes, and reduces to the above degree-based approximations in special cases~\cite{Faqeeh15}.

As these theories assume the network is locally treelike, they are prone to errors in clustered networks which have an appreciable density of short loops \cite{Melnik11,Faqeeh15}. However, on some real-world clustered networks these theories still perform well, and in some other cases the inaccurate predictions of these theories are shown to be only partly caused by the presence of short loops \cite{Faqeeh15}. This indicates the presence of an unexplained source of error and possibly a phenomenon not captured by the theories.

In this paper, we show the appearance of coexisting percolating clusters (CPCs) in certain networks, and demonstrate that this phenomenon causes significant errors in the aforementioned theories. We show that CPCs appear in modular networks with limited mixing, i.e., networks with a sufficiently small (limited) number of interlinks between modules. We verify these observations by developing the modular message passing (MMP) theory which takes into account the presence of independent CPCs. We show that the MMP theory provides accurate predictions on treelike modular networks with limited mixing and also improves over the predictions of MP on several real-world clustered networks.

\begin{figure}[t!] \centering
\captionsetup{farskip=0pt}
\subfloat{\includegraphics[width=.49\columnwidth]{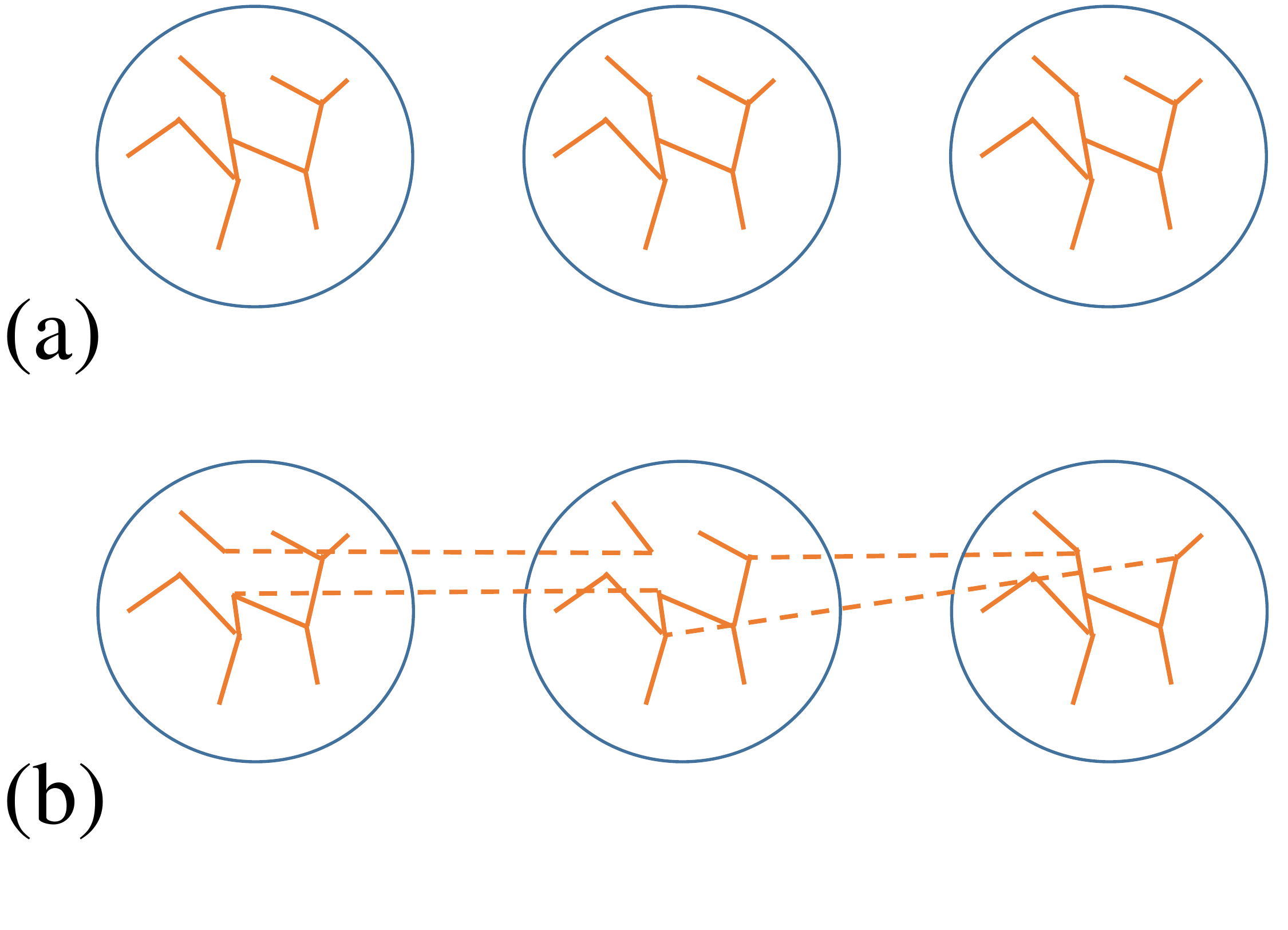}\label{f1a}}
\subfloat{\label{f1b}}
\hfill
\subfloat{\includegraphics[width=.5\columnwidth]{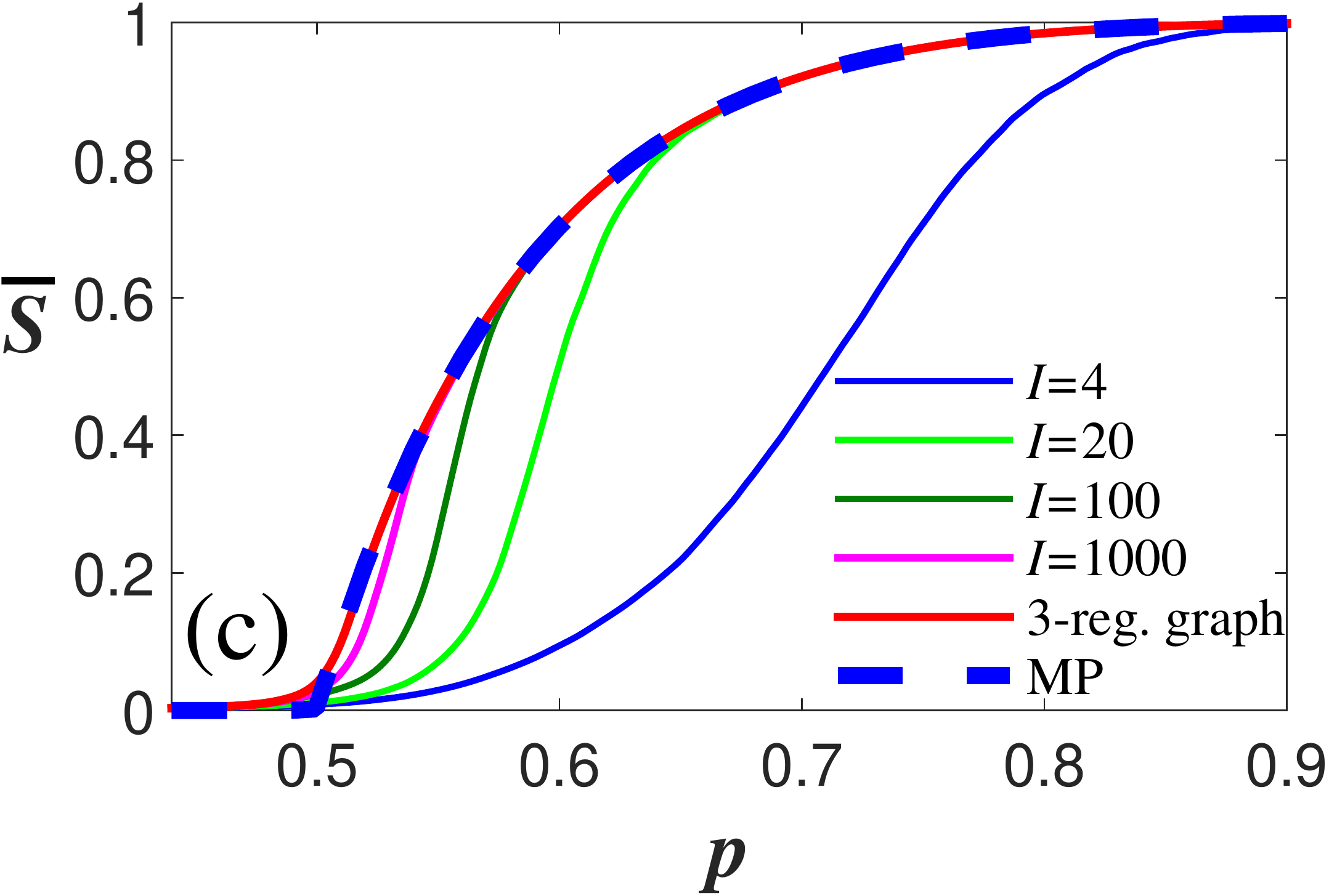}\label{f1c}}\\[-1.5ex]
\captionsetup{farskip=10pt}
\subfloat{\includegraphics[width=.5\columnwidth]{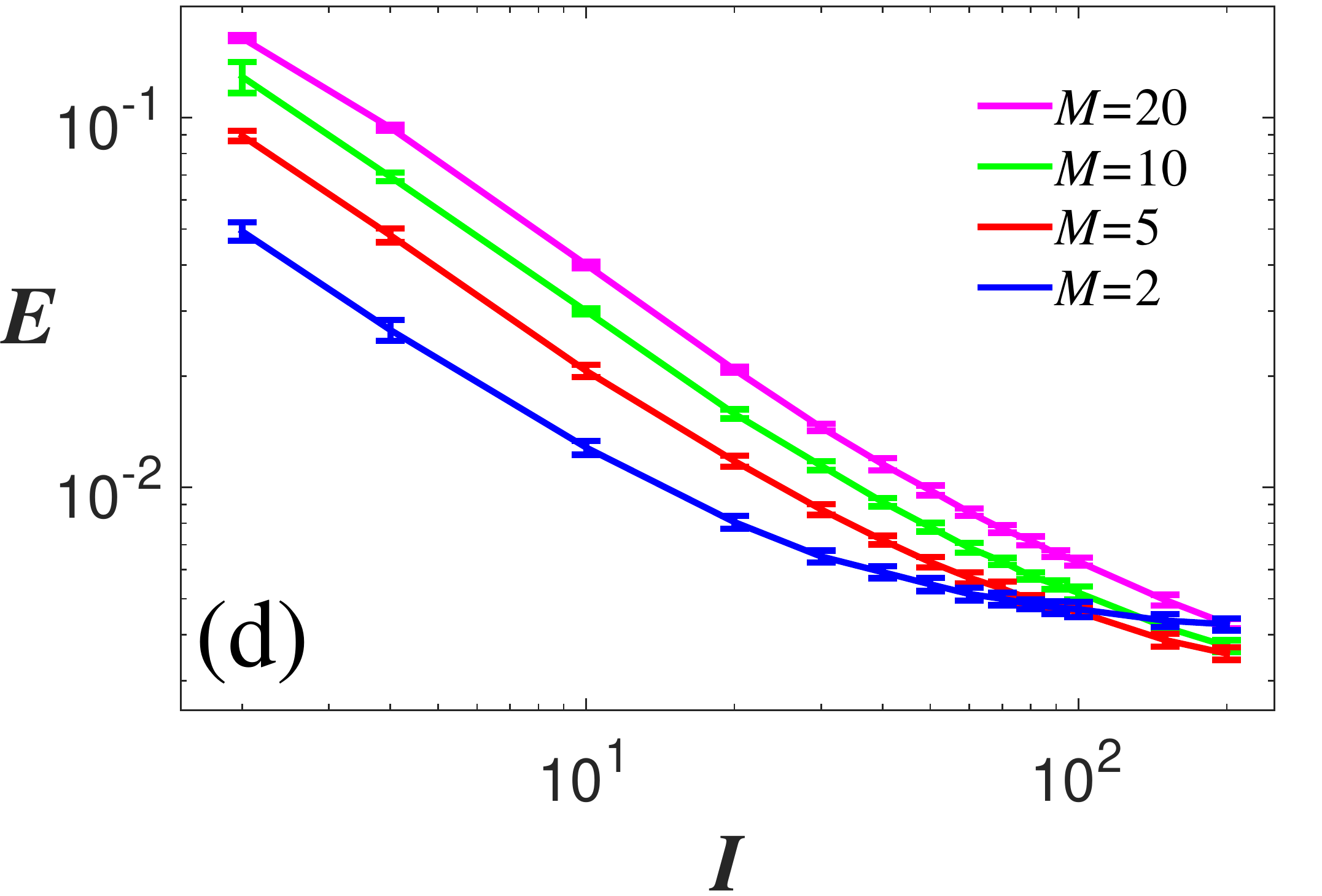}\label{f1d}}
\hfill
\subfloat{\includegraphics[width=.5\columnwidth]{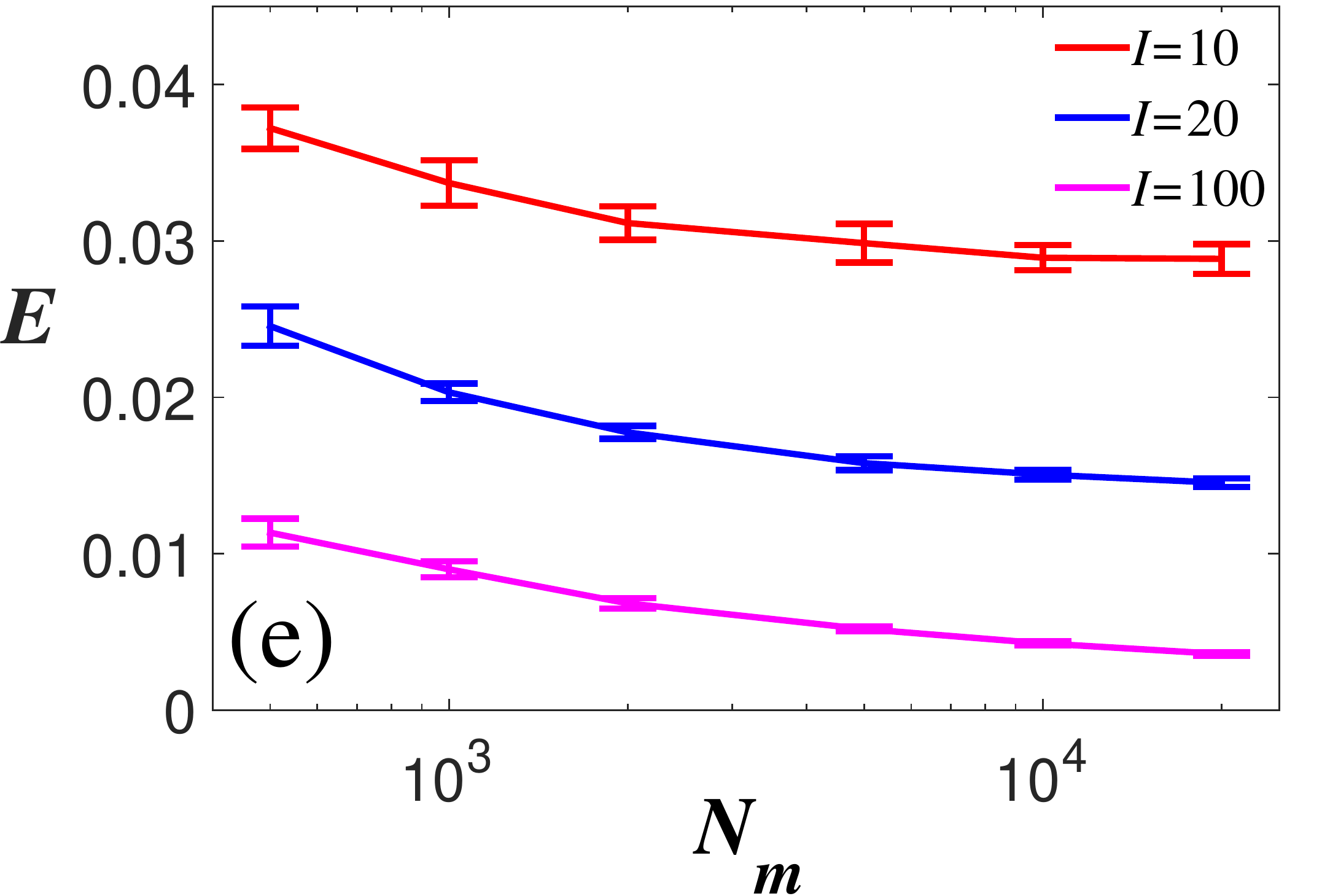}\label{f1e}}
\caption{(Color online) (a) A schematic of $M=3$ identical copies of a graph, (b) that are connected in a chain to construct an SLN with $M$ modules, each pair of modules connected through $I$ interlinks. (c) Bond percolation results on SLNs constructed from a 3-regular graph of size $N_m=1000$ for different $I$. (d) The mean absolute error $E$ versus $I$ and (e) versus $N_m$ for the MP theory on such 3-regular SLNs. In (c) $M=50$, in (d) $N_m=5000$, and in (e) $M=10$. }
\end{figure}

We begin by introducing sausage-like networks (SLNs), a simple ensemble of random networks that demonstrates the appearance of CPCs. To create an SLN we first pick a graph with size $N_m$, which can be any connected undirected unweighted graph. Then we make $M$ identical copies of that graph (Fig.~\ref{f1a}) which will become modules in the SLN. We assign to each of these modules a unique label $m\in\{1,2,\ldots,M\}$ and connect each pair of modules with consecutive labels $m$ and $m+1$ by $I$ links. To do so, exactly $I/2$ links are selected randomly from module $m$. For each selected link $i_m\mbox{---}j_m$ we consider its copy $i_{m+1}\mbox{---}j_{m+1}$ in module $m+1$, and rewire these two links to create two new links $i_{m}\mbox{---}j_{m+1}$ and $i_{m+1}\mbox{---}j_{m}$ instead. The resulting SLN is comprised of a chain of modules (Fig.~\ref{f1b}), each pair of consecutive modules connected with exactly $I$ interlinks. Moreover, an SLN has a degree distribution and degree-degree correlations between and beyond the nearest neighbors identical to those of the original graph \cite{Faqeeh15}. Similarly, one constructs SLNs from non-identical modules by rewiring links that are not copies but randomly selected from each module.

In Figs.~\ref{f1c}-\ref{f1e}, we illustrate the bond percolation results on SLNs constructed from a 3-regular graph. The results from the MP theory are obtained using the Eqs.~(1) and (2) of Ref.~\cite{Faqeeh15} (originally due to \cite{karrer_PRL14}):
\begin{eqnarray}
  u_{ij}=1&-&p+p\prod_{k\neq i}{A}_{jk}u_{jk},\\
  \mathit{\overline{S}}=\frac{1}{N}\sum_{i=1}^Ns_i&=&1-\frac{1}{N}\sum_{i=1}^N\prod_j{A}_{ij}u_{ij},\label{eq0b}
\end{eqnarray}
where $p$ is the occupation probability, $u_{ij}$ is the probability that node $i$ is not connected to the network PC via its link to $j$, $A$ is the network adjacency matrix, $N$ is the size of the network, and $s_i=1-\prod_j{A}_{ij}u_{ij}$ is the probability that node $i$ is in the PC.

Numerical simulations show that the behavior of $\mathit{\overline{S}}$ in the SLNs depends substantially on the number of interlinks $I$ and the number of modules $M$. Surprisingly, the result of the MP theory for the SLNs is independent of $I$ and $M$ and coincides with its prediction for a single 3-regular graph (Fig.~\ref{f1c}); it is worth mentioning that in such SLNs the result of MP theory is the same as the results of any of its degree-based reductions mentioned above. The numerical results, on the other hand, deviate from the theoretical prediction as the number of interlinks $I$ is decreased (Fig.~\ref{f1c}). We quantify the difference between the theoretical and numerical results by calculating the mean absolute error between the two: $E=1/R\sum_{j=1}^R|\mathit{\overline{S}}_{theo}(p_j)-\mathit{\overline{S}}_{num}(p_j)|$, where the sum is over $R=100$ equally spaced occupation probabilities $p_j=j/R$.

Figure \ref{f1d} shows that the error increases dramatically for lower number of interlinks $I$; on the other hand the error increases only slightly for smaller module sizes $N_m$ (Fig.~\ref{f1e}). Figure~\ref{f1e} highlights also that the error increases for lower $I$ rather than for lower ratio of $I$ to all edges.

\begin{figure}[t!] \centering
\captionsetup{farskip=0pt}
\subfloat{\includegraphics[width=.5\columnwidth]{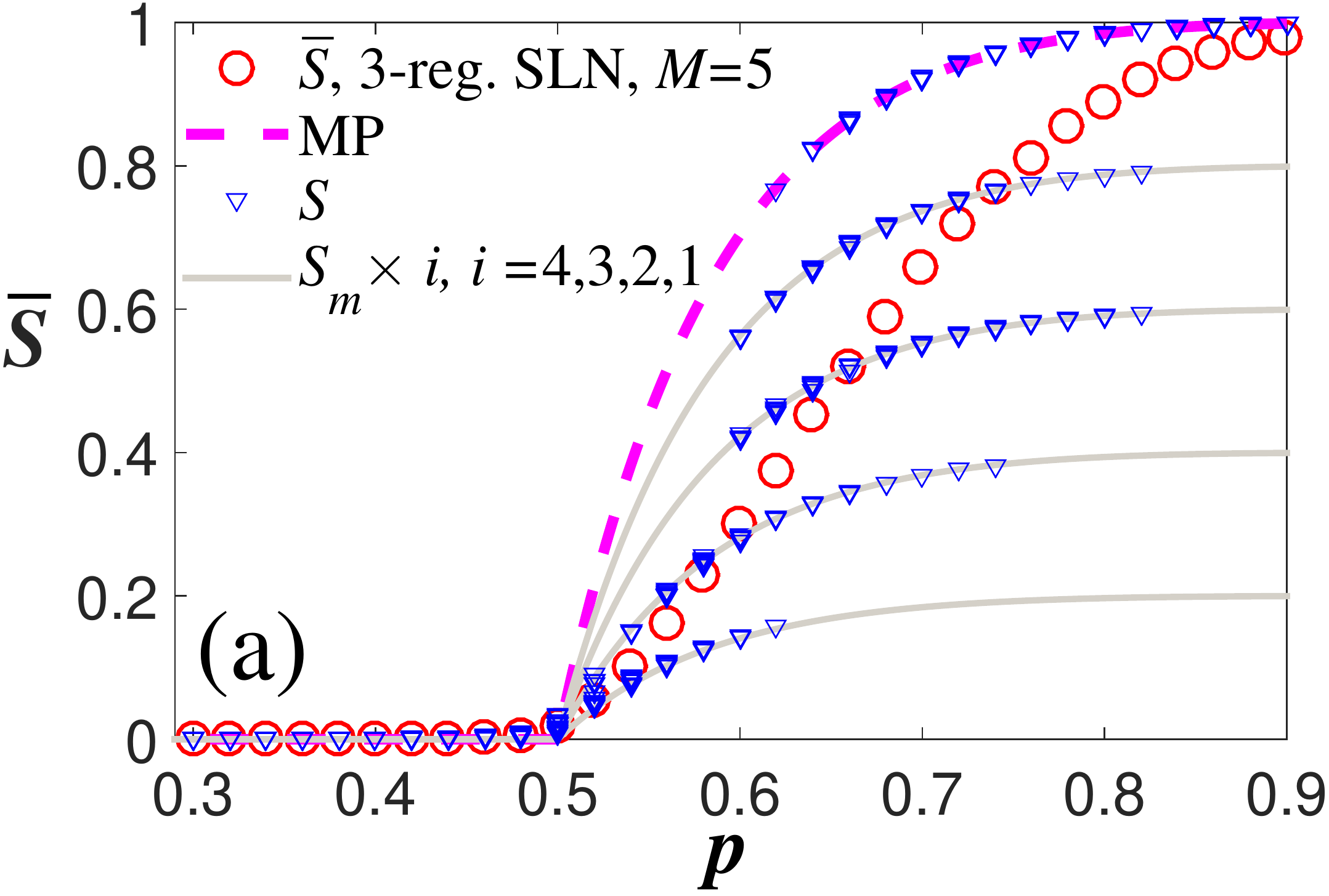}\label{f2a}}
\hfill
\subfloat{\includegraphics[width=.5\columnwidth]{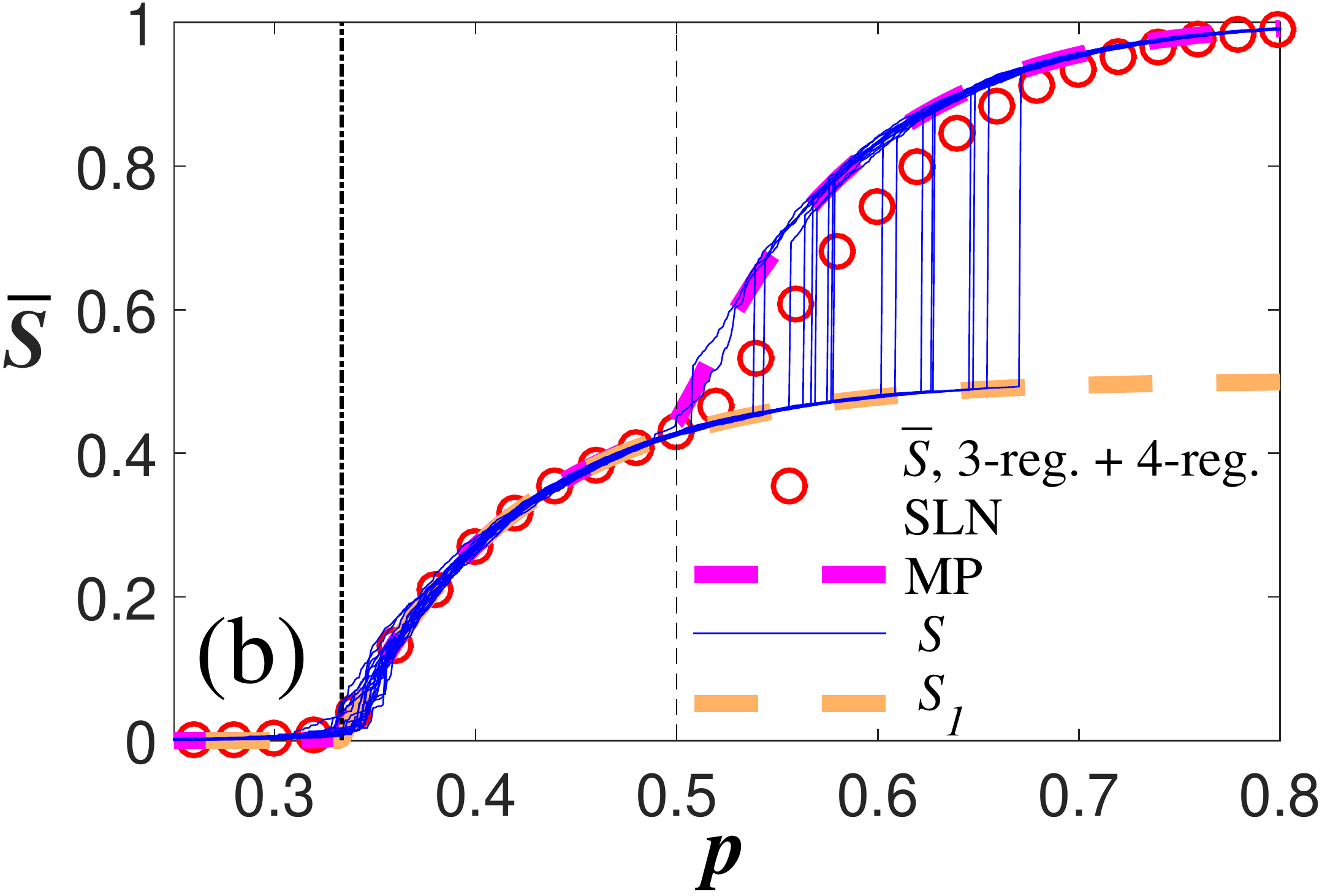}\label{f2b}}
\\[-2ex]   \captionsetup{farskip=10pt}
\subfloat{\includegraphics[width=.5\columnwidth]{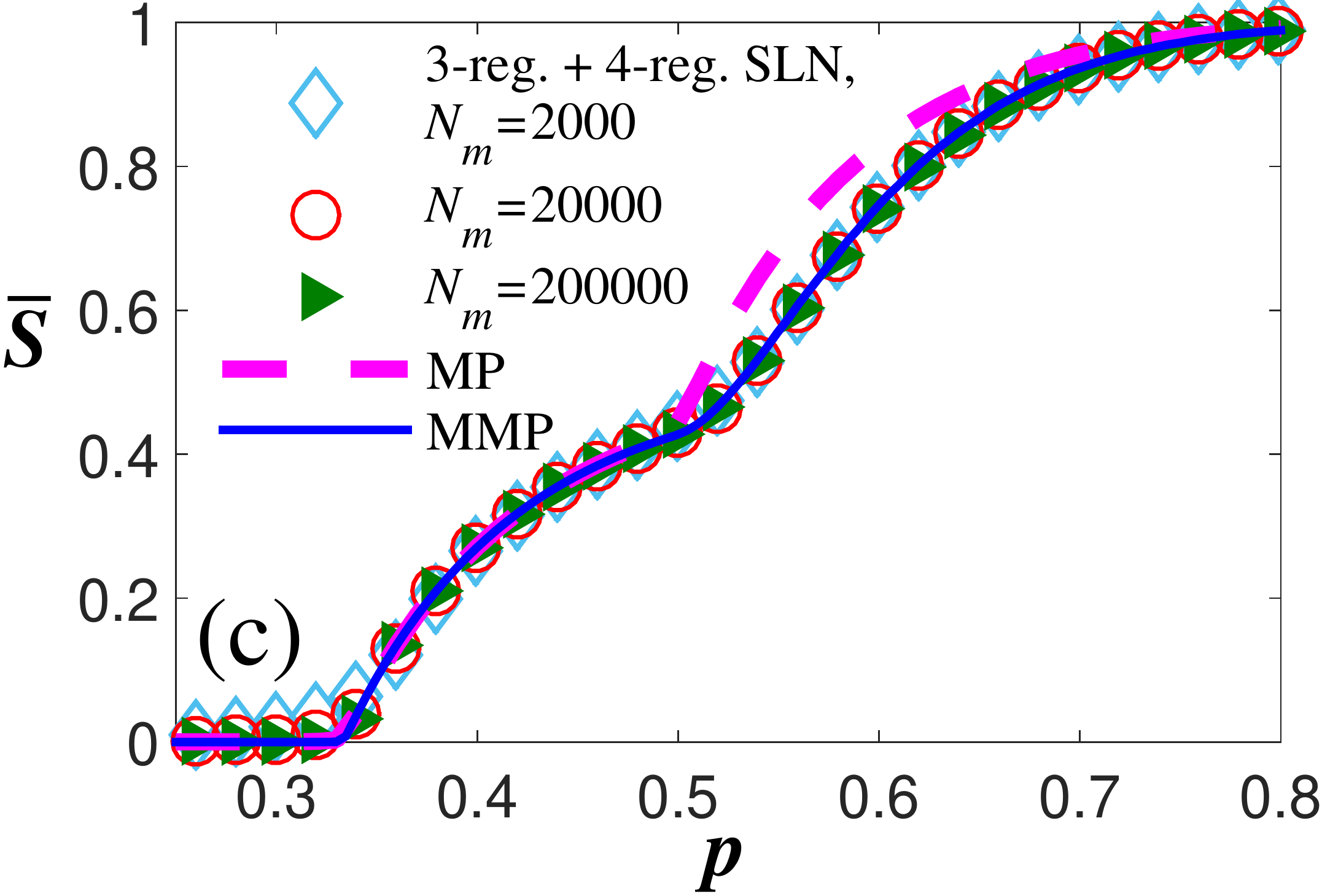}\label{f2c}}
\hfill
\subfloat{\includegraphics[width=.5\columnwidth]{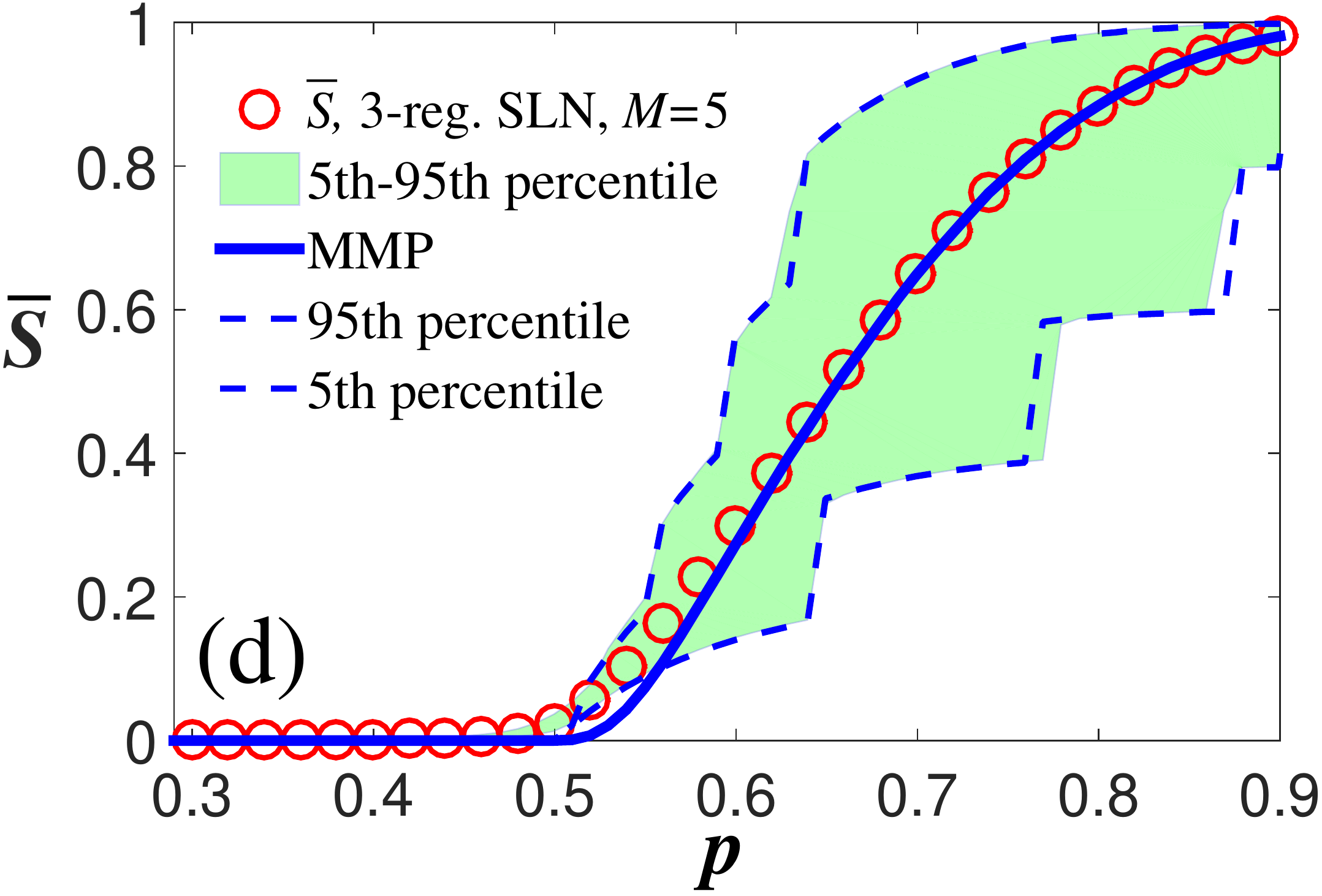}\label{f2d}}
\caption{(Color online) (a) Percolation on an SLN with $M=5$ and $N_m=20000$. For each $p$, $S$ is shown separately for 20 single realizations of percolation (triangles). $S$ for single realizations does not match $\mathit{\overline{S}}$ (circles) averaged over 500 realizations, nevertheless it coinsides with one of the 5 possible values denoted by the dashed line (MP theory) or the solid lines. (b) Percolation on an SLN with a 3-regular and a 4-regular module, $I=4$, and $N_m=20000$. Blue lines represent the results from 20 single realizations of the Newman-Ziff algorithm. The vertical line on the left (right) denotes the percolation threshold of a 4-regular (3-regular) graph. (c) The MMP prediction for $\mathit{\overline{S}}$ on the SLN of panel (b) matches the numerical results on this SLN (circles) as well as on similar SLNs with identical $I$ and any sufficiently large $N_m$. (d) Predictions for the SLN of panel (a). The 5th and 95th percentiles of $S$ are indicated by the green shade (numerical results) and by the dashed lines (MMP theory).
}
\end{figure}

To understand these observations, it is necessary to inspect also the numerical results for $S$, the fractional size of
the network PC for single realizations of the bond percolation process. Figure \ref{f2a} shows the results for a 3-regular SLN with $M=5$ modules; the MP theory overestimates $\mathit{\overline{S}}$, which is the expected (average) value of $S$. For a sufficiently large network, we normally expect that $S$ fluctuates slightly around $\mathit{\overline{S}}$. Surprisingly, we observe in Fig.~\ref{f2a} that, for a fixed value of $p$ (in a certain interval), $S$ can take one of several possible values which can be significantly different from $\mathit{\overline{S}}$. The different possible values for $S$ can be explained as follows. Let us denote by $S_m$ the fraction of nodes in module $m$ that are in the network PC predicted by the theory. Hence, for the MP theory $S_m=\frac{1}{N}\sum_{i\in\hat{m}}s_i$, where $\hat{m}$ denotes the set of nodes located in module $m$, and Eq.~\eqref{eq0b} becomes:
\begin{equation}\label{eq0c}
  \mathit{\overline{S}}=\sum_{m=1}^MS_m.
\end{equation}
In Fig.~\ref{f2a}, we can observe that $\mathit{\overline{S}}$ predicted by MP matches the largest possible value of $S$. The lowest possible value of $S$, on the other hand, coincides with $S_m$ of only one module. The next 3 larger  possible values of $S$ coincide with the sum of $S_m$ for 2, 3, and 4 modules respectively (Fig.~\ref{f2a}).

Consider another example of an SLN consisting of one 3-regular module and one 4-regular module each having the same number of nodes $N_m$, and let us denote by $S_1$ and $S_2$ the $S_m$ of the 4-regular and 3-regular graph respectively. For this SLN, $\mathit{\overline{S}}$ from numerics and from the MP theory both match $S_1$ up to $p_{3reg}=0.5$, the percolation threshold of a 3-regular graph. Above this value, the MP prediction deviates from the numerical result (Fig.~\ref{f2b}). This deviation can be better understood by looking at single realizations of the Newman-Ziff algorithm \cite{Newman01e}, where starting with no occupied links, we occupy links one by one in random order. As the network is large, $p$ is approximately equal to the fraction of occupied links. In Fig.~\ref{f2b}, we can observe that up to $p_{3reg}$, single realization values of $S$ match the value of $\mathit{\overline{S}}$. However, above $p_{3reg}$, while the predicted $\mathit{\overline{S}}$ is $S_1+S_2$, $S$ will remain equal to $S_1$ until some larger value of $p$ and then suddenly jump to $S_1+S_2$.

This implies that although $S_2$ is finite for $p\geqslant p_{3reg}$, it does not represent the probability a node in module 2 belongs to the network PC, and accordingly $S_1+S_2$ is not the expected size $\mathit{\overline{S}}$ of the network PC. In fact, the nodes in module 2 are part of a PC with size $S_2$ and those in module 1 are part of another PC with size $S_1$. In a single realization these two PCs may be unconnected. Then, when more links are occupied one at a time, at a value of $p$ with $p\geqslant p_{3reg}$ the two PCs become suddenly connected, and the size of the network PC changes abruptly from $S_1$ to $S_1+S_2$. We refer to such PCs, which exist independently of each other in the network but may only be connected with a finite probability,  as coexisting percolating clusters (CPCs).

Next, we develop the modular message passing (MMP) theory to describe and verify the phenomenon of CPCs appearing in networks. The two main assumptions are (i) modules can percolate independently (hence, the appearance of independent CPCs), and (ii) PCs of neighboring modules are connected with  probability $\eta<1$. This is a new concept different from the common assumption (see for example \cite{Newman10,Melnik_chaos14,colomer2014double}) that there exists only one monolithic PC in a network; here we show that in networks with limited mixing, the network PC is polylithic, i.e., constituted by CPCs that are connected together. In such networks, the CPCs emerge independently inside the internally well connected modules, but they still may not be connected to each other due to the small number of interlinks between the modules. For $p<1$, interlinks may be unoccupied; the CPCs of two neighboring modules are connected if and only if they share at least one occupied interlink that connects nodes from the CPCs of the two modules.

For SLNs we assume that each boundary node (a node with links to other modules) has exactly one link to a neighboring module. Then for an SLN with two modules, our MMP theory is comprised of two simple equations. First we calculate $\eta_{mn}$, the probability that the CPCs of the two modules $m$ and $n$ are connected:
\begin{equation}\label{eq1}\centering
  \eta_{mn}=1-\left(1-pv_mv_n\right)^I.
\end{equation}
Here $p$ is the occupation probability and $v_m$ is the probability that a boundary node in $m$ is part of the CPC of $m$.
Then $1-pv_mv_n$ is the probability that the two CPCs are not connected via an interlink; raising this term to the power of $I$ gives the probability that they are not connected via any of the $I$ interlinks.
Equation~\eqref{eq1} is independent of $N_m$, hence in the thermodynamic limit ($N\to\infty$), if $I$ is fixed to a sufficiently small number, (independent) CPCs emerge, connected with a probability $\eta_{mn}(p)<1$. Whereas for large $I$, since $\eta_{mn}\to 1$ the two CPCs are connected with high probability, leading to a monolithic PC.
For the SLNs described above, $v_m$ and $v_n$ can be simply calculated using the $p_k$ theory (see Sec.~S.1. of the Supplemental Material (S.M.) \cite{SM}). For an SLN with two modules, the size of network PC is then:
\begin{equation}\label{eq2}
  \mathit{\overline{S}}=\eta_{12}\left(S_1+S_2\right)+\left(1-\eta_{12}\right)S_1,
\end{equation}
where $S_1$ and $S_2$ are the fractional sizes of respectively the larger and the smaller CPC of the SLN. Figure~\ref{f2c} shows that the prediction of the MMP theory (Eqs.~\eqref{eq1}-\eqref{eq2}) matches perfectly the numerical result for the SLN of Fig.~\ref{f2b}.

For SLNs with more than two modules Eq.~\eqref{eq1} can still be used to calculate $\eta_{mn}$ for each pair of modules $m$ and $n$. In the case when boundary nodes have more than one interlink or when the connection pattern of the modules can not be well approximated using the $p_k$ theory, Eq.~\eqref{eq1} should be extended to include more information on the network structure. We can use the full information on the adjacency of individual nodes to write a general formula for the connection probabilities $\eta_{mn}$ between CPCs in a treelike network:
\begin{equation}\label{eq3}
\eta_{mn}=1-\prod_{i=1}^I\left[1-\left(1-\prod_{j\in\mathcal{A}_m(i)}u_{ij}\right)\left(1-\prod_{k\in\mathcal{A}_n(i)}u_{ik}\right)\right],
\end{equation}
where $i$ is a boundary node of module $m$ and $\mathcal{A}_m(i)$  denotes the set of neighbors of $i$ in modules $m$. Here, $u_{ij}$ and $u_{ik}$ are the probabilities that $i$ is not connected to the CPC of, respectively, modules $m$ and $n$ via its links to nodes inside each of those modules. Hence, within the first (second) set of parentheses in Eq.~\eqref{eq3} is the probability that $i$ is in the CPC of $m$ (the CPC of $n$), and in the square brackets we have the probability that the two CPCs are connected via the interlinks of $i$. Therefore, $\eta_{mn}$ is the probability that the two CPCs are connected via any of their $I$ interlinks.

To calculate $\mathit{\overline{S}}$ for networks with more than two modules (and consequently more than two CPCs), Eq.~\eqref{eq2} should be extended as well. If a networks contains several CPCs then in a single realization of percolation, different CPCs (with sizes $S_m$) are connected together with some probability, creating larger polylithic PCs. The polylithic PC $l$ has a size $S^{\rm(pol)}_{l}=\sum_{m\in\hat{l}}S_m$, where $\hat{l}$ denotes the set of CPCs that constitute $l$.
Then in a single realization at a fixed value of $p$, the size $S$ of the network PC is $\max_l\left(S^{\rm(pol)}_{l}\right)$, i.e., the size of the largest polylithic PC. Then the expected size of the network PC is
\begin{equation}\label{eq4}
  \mathit{\overline{S}}=\sum_{S}P_{\left(S\right)}{S},
\end{equation}
where $P_{\left({S}\right)}$ is the probability that in a single realization the size of the largest polylithic PC is ${S}$. To calculate $P_{\left({S}\right)}$, for each $p$ we first calculate $S_m(p)$ values using the MP theory or an appropriate degree-based reduction of MP. Then we assume a meta-network which is comprised of meta-nodes; each meta-node $m$ represents a module of the original network and has a weight $S_m(p)$. In each realization of the meta-network for a fixed value of $p$, each pair of meta-nodes $m$ and $n$ are connected with probability $\eta_{mn}(p)$. For sufficiently large number of meta-network realizations, we calculate $P_{\left({S}\right)}$ using the Newman-Ziff algorithm \cite{Newman01e} with the following modifications: (i) each link is added with probability $\eta_{mn}(p)$, and (ii) the size of a cluster $l$, comprised of connected meta-nodes, is the sum of the weights of meta-nodes it includes, i.e., $S^{\rm(pol)}_{l}$. These calculations are performed very quickly, as the number of the modules is usually much smaller than the number of nodes.

Figure~\ref{f2d} illustrates that, using Eqs.~\eqref{eq3}-\eqref{eq4}, the MMP theory performs very well for the SLN of Fig.~\ref{f2a}. As shown in Figs.~\ref{f2a} and \ref{f2b} in the presence of CPCs the values of $S$ for single realizations can deviate considerably from the expected value $\mathit{\overline{S}}$. In the MMP theory, the variability of $S$ is determined by $P_{(S)}$ defined above. The percentiles calculated from $P_{(S)}$ match very well the numerical values in SLNs (see Fig.~\ref{f2d} for example), which confirms that the high variability of $S$ originates from the presence of CPCs that suddenly merge together. Similar results (not shown) are obtained for SLNs with different values of $I$, $M$, $N_m$, and also for SLNs constructed from modules with a heterogenous structure, i.e., with a power-law degree distribution.

\begin{figure}[t!] \centering
\captionsetup{farskip=0pt}
\subfloat{\includegraphics[width=.5\columnwidth]{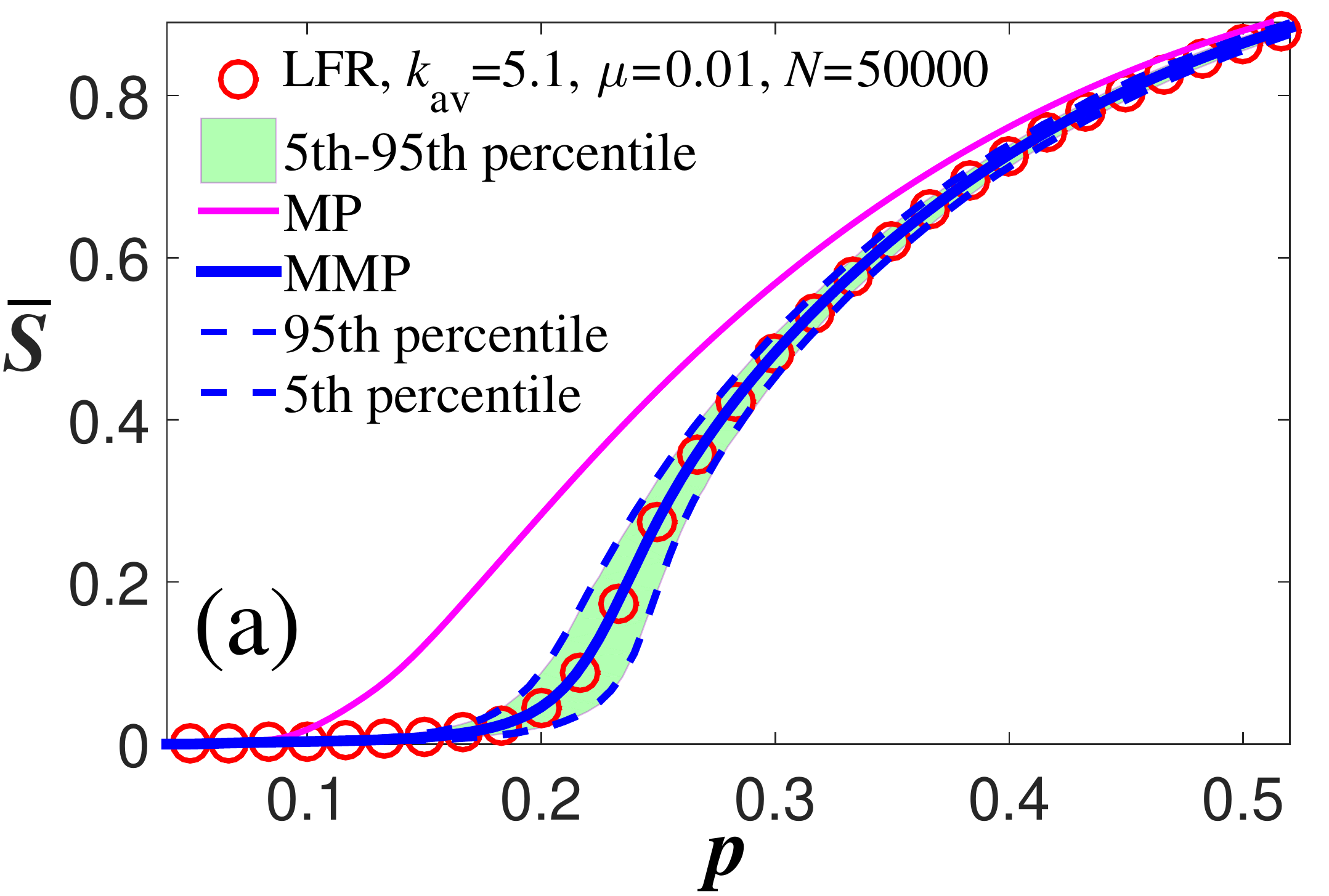}\label{f4a}}
\hfill
\subfloat{\includegraphics[width=.5\columnwidth]{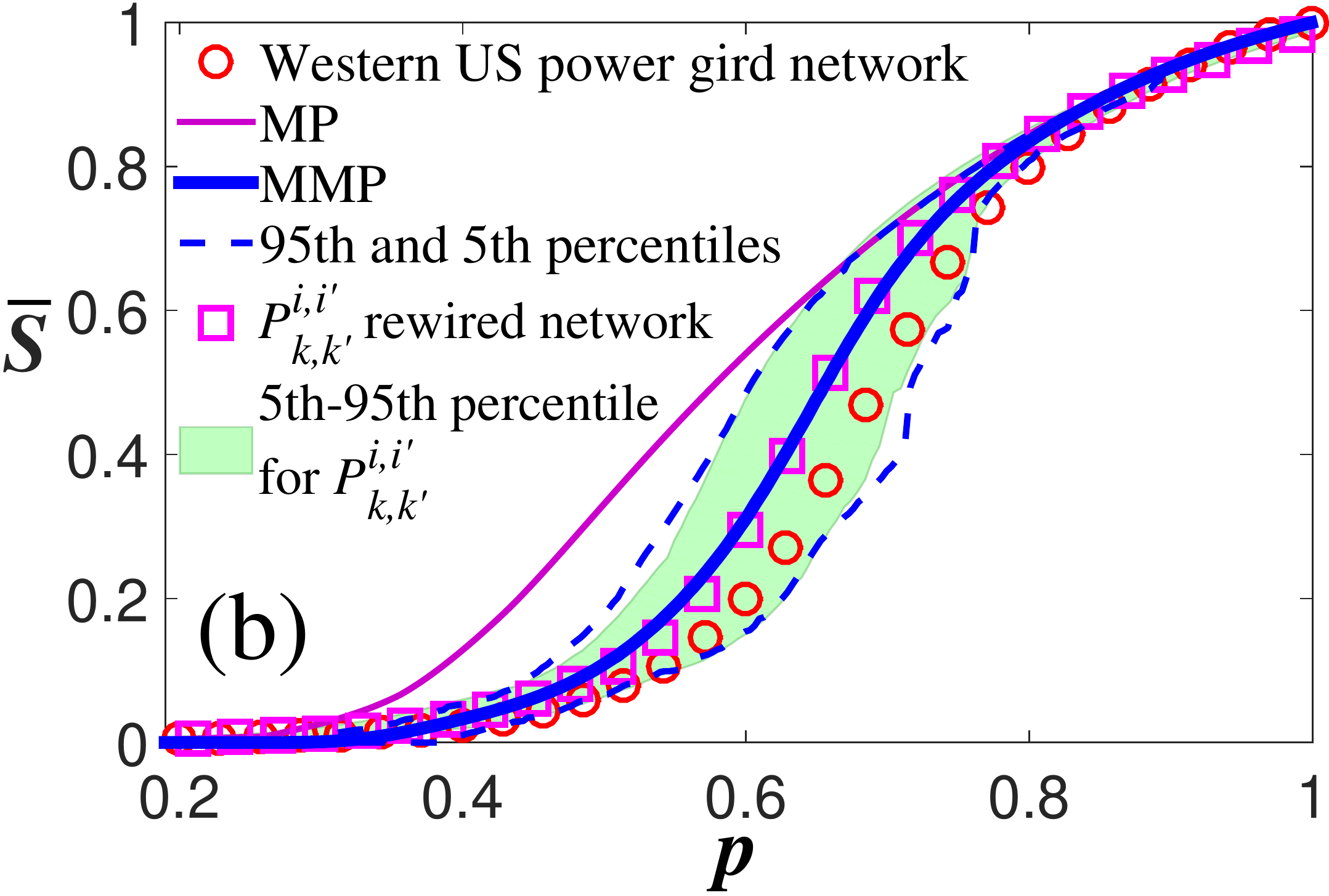}\label{f4b}}
\caption{(Color online) (a) Bond percolation results on an LFR network with $N$ nodes, degree distribution $P(k)\propto k^{~-3}$, community size distribution $P(N_m)\propto N_m^{~-1}$, average degree $k_{\rm av}$, the mixing parameter $\mu$, and minimum and maximum community sizes of respectively 50 and 1000. (b) Results for the western US power grid network. For the LFR network, MMP predicts accurately $\mathit{\overline{S}}$ and also percentiles for the distribution of $S$; for the power grid its predictions match the results for the $P_{k,k'}^{i,i'}$ rewired version of this network.}
\end{figure}

We also provide results for LFR benchmark networks \cite{Lancichinetti08}, as an example of modular networks with heterogeneous structure. In LFR networks, all pairs of modules can be connected without any restriction (as apposed to SLNs), and the node degrees and community (module) sizes have a power-law distribution. On LFR networks with low mixing between modules, the numerical results do not match the MP predictions, whereas MMP method provides accurate predictions (e.g., see Fig.~\ref{f4a}).
As mentioned before, MMP assumes that modules have (independent) CPCs. This assumption holds only for sufficiently large modules. Nevertheless, if the size of the modules is as small as $N_m^{\rm(min)}\simeq50$, MMP still performs well (see S.M., Sec.~S.2. \cite{SM}).

We can show that on several real-world networks the bond percolation results are affected by the emergence of CPCs. To do so, we first identify the best representations of network modular structure using a multiresolution community detection method (e.g., see Refs.~\cite{Ronhovde09,blondel08,Faqeeh12,Rosvall08}). Then, we choose the representation that maximises the modularity $Q$ \cite{Newman04_pre} and minimizes $M_s$ the number of modules with sizes smaller than $N_m^{\rm(min)}=50$, i.e., the representation with maximum $\frac{(M-M_s)}{M}Q$. Figure~\ref{f4b} shows that MMP improves significantly over the MP prediction of $\mathit{\overline{S}}$ on the western United States power grid network \cite{Melnik11,Watts98}, and also provides a prediction for the variability of $S$ according to the percentiles of $P_{(S)}$. The results of MMP on the power grid network, match the numerics for the $P_{k,k'}^{i,i'}$ rewired ~\cite{Melnik_chaos14} version of this network in which the links are rewired inside each module; $P_{k,k'}^{i,i'}$ rewiring preserves the modular structure and degree-degree correlations but effectively destroys the short loops. This shows that MMP provides a highly accurate prediction in the absence of short loops and when the modular structure is identified accurately. Similar results are shown in S.M. for several other examples of real-world networks \cite{SM}.

In summary, we demonstrated that CPCs can emerge in ensembles of random networks and in real-world networks, when the network modules are connected via a small number of interlinks. Moreover, we showed that CPCs are an important source of error in the theories for bond percolation and proposed the MMP theory that accurately captures the impact of CPCs on percolation results. An important implication of the appearance of CPCs is the uncertainty they cause in determining the network robustness: when the CPCs emerge, the size ${S}$ of the network PC can be highly variable (as shown by the percentiles). This implies the prominent role of interlinks in network robustness, even in the absence of module-based targeted attacks \cite{daCunha15}, and subject to only random failures. Another implication is that the eventual size of an epidemic spread in the SIR model~\cite{Newman02b,Karrer10} may not be best represented by $\mathit{\overline{S}}$, which is the expected size of the largest polylithic PC, since even CPCs not in the largest polylithic PC represent (independent) outbreaks of comparable sizes located in different modules. Hence, the total size of an epidemic outbreak may better be represented by the sum of the sizes of all CPCs.

We would like to thank Mari\'an Bogu\~n\'a, Mel Devine, and David O'Sullivan for valuable comments.~We acknowledge funding from Science Foundation Ireland (under programmes 11/PI/1026 and 12/IA/1683) and from the European Commission FET-Proactive project PLEXMATH (FP7-ICT-2011-8; Grant No. 317614).
\bibliography{networks}

\end{document}